%% file: sample631.tex
\documentclass[]{aastex631}

\usepackage{amsmath}
\graphicspath{{./}{figures/}}

\begin{document}

\title{Calculating Occultation Light Curves using Wavelets: Exponential Atmospheres and the Constraints of Static Stability}

    \author[0000-0002-7547-3967]{Leslie A. Young}
    \affiliation{Southwest Research Institute \\
    1050 Walnut St., Suite 300 \\
    Boulder, CO 80540, USA}
    
\author[0000-0003-0000-0572]{Michael J. Person}
\affiliation{MIT, Department of Earth, Atmospheric, and Planetary Sciences \\
77 Massachusetts Ave. \\
Cambridge, MA 02139, USA}



\begin{abstract}

The signatures of waves are seen during many high-quality ground-based refractive stellar occultations by solar system atmospheres. We present a new forward-modeling technique for ground-based stellar occultations based on wavelet decomposition.  If profiles of refractivity are written as the product of an exponential and a wavelet decomposition, then we can analytically write the profiles of the bending angles and the bending angle derivatives that are needed to calculate occultation light curves.
Requiring that the atmosphere is statically stable places limits on the amplitudes of atmospheric waves and their effect on the observed light curve.

\end{abstract}



\section{Introduction} \label{sec:intro}

Atmospheric waves are important to the momentum or energy budgets of most of the upper atmospheres in our solar system.  While some waves have large vertical wavelengths that are easily resolved even by relatively low-resolution methods of remote sensing such as infrared sounding (e.g. \citealt{HinsonWilson2023Icar..39415420H}),
stellar occultations are one of the few methods of remote sensing that can resolve atmospheric waves with sub-scale-height vertical wavelengths \citep{FrenchGierasch1974JAtS...31.1707F, Roques+1994A&A...288..985R, Sicardy2022CRPHYS....23.213S, CoorayElliot2003ApJ...587L.121C}. Stellar occultations reveal light curves, or the change in flux vs.\ time as a star sets behind or emerges from behind a planetary atmosphere. The abscissa of such a curve is often expressed in terms of the shadow radius, $y$, or the distance between the observer and the center of the shadow, on a plane passing through the observer and perpendicular to the unit vector to the star (see e.g. \cite{Sicardy2022CRPHYS....23.213S}).

\begin{figure}  [ht]
    \centering
    \includegraphics[width=1.0\linewidth]{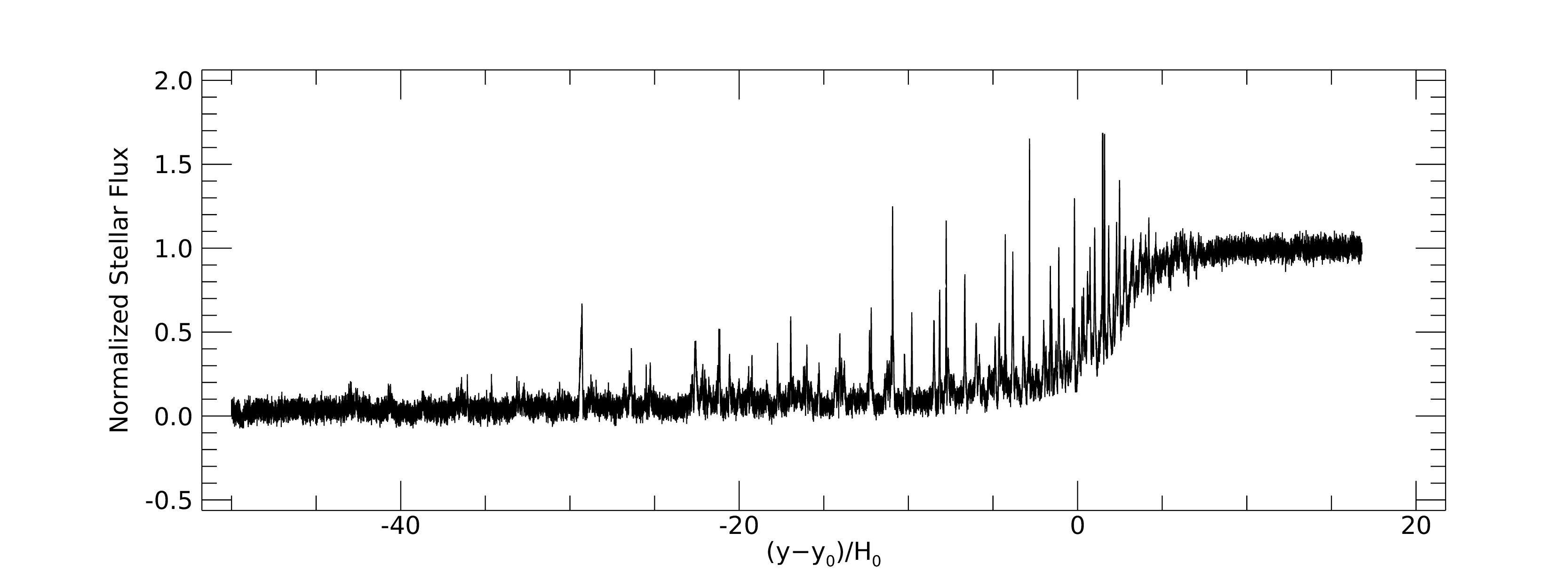}
    \caption{The Uranus occultation of 1995 Sept 9 from SAAO.  Here we plot the immersion in normalized stellar flux vs.\ distance in the shadow plane, relative to half light and scaled by the atmospheric scale height. Altitude increases to the right, and, for this immersion light curve, time increases to the left. This light curve shows many of the characteristics discussed in this paper such as multiple examples of resolved spikes, some that reach higher than the unocculted flux level, and some that are observed out to 45 scale heights with wider spacing and decreased maximum flux.}
    \label{fig:ur}
\end{figure}

The Uranus occultation (Figure \ref{fig:ur}) of 1995 Sept 9 \citep{French+2023Icar..39515474F, Saunders+2024PSJ.....5..247S} shows some of the typical characteristics of many occultations by giant planets or Titan. (1) The light curve is dominated by narrow spikes from waves or turbulence. (2) These spikes often exceed the unocculted stellar flux. (3) The spikes are narrow but often resolved, with broader troughs between the spikes. (4) The spikes are highest when the average flux is near halflight, and decrease in the tail. (5) In the tail of the light curve, the time between spikes increases.  The reduction of occultations such as this have generally followed one of three methods: inversion of the data,  statistical characterization of the spikes, or parameterized forward modeling.  

The inversion method uses an Abel transform to calculate a refractivity profile from an observed light curve \citep{WassermanVeverka1973Icar...20..322W, French+1978Icar...33..186F,Elliot+2003AJ....126.1041E}. At its best, this method achieves the highest possible spatial resolution, since each observed value of flux translates into a derived refractivity in the atmosphere. However, this process requires very high signal-to-noise ratios on the raw data, and cannot handle negative fluxes, so data is often pre-averaged or smoothed. This method assumes geometric optics, no extinction, no ray crossing (where the starlight is overfocused by the atmosphere), and that the star is a point source. Furthermore, the statistical significance of the derived temperatures is difficult to characterize \citep{Harrington+2010ApJ...716..404H}. 

Another method concentrates on characterizing the statistical properties of the fluctuations in the refractivity and the phase screen from the statistical deviations of an observed light curve, compared to that produced by an isothermal atmosphere \citep{Hubbard+1978Icar...34..374H}. This directly addresses the issue of describing the putative waves or turbulence, making predictions on the spacing of spikes deep in the tail or the variation of flux amplitude \citep{NarayanHubbard1988ApJ...325..503N}. However, it does not attempt to derive profiles of refractivity or temperature. 

In parameterized forward modeling \citep{WassermanVeverka1973Icar...20..322W, ChamberlainElliot1997PASP..109.1170C}, the atmosphere is described as a model with some small number of parameters. The simplest such model is an isothermal atmosphere, which depends on only two parameters, such as the scale height and the refractivity at a reference height. More complex functions can include thermal gradients \citep{ElliotYoung1992AJ....103..991E}, more elaborate ad hoc functions \citep{Yelle+1996JGR...101.2149Y}, or be based on sophisticated atmospheric models \citep{Zalucha2011Icar..214..685Z}. With forward modeling, the integrals of refractivity are calculated, including effects such as ray crossing, wave optics, and the finite angular size of the occulted star. The model light curves can be compared directly to observations, with best-fit parameters and formal errors typically calculated via estimation methods such as least-squares or Markov Chain Monte Carlo. The required line-of-sight integrals can be slow, which is rarely a limitation unless many light curves need to be calculated (e.g., for Monte Carlo noise analysis). A more severe limitation is that the models are rarely flexible enough to account for the small-scale structure at vertical wavelengths much smaller than a scale height.

A newer approach describes an atmosphere as the product of a mean profile and a perturbation, and generates a light curve from the Fourier decomposition of the perturbation \citep{Young2009AJ....137.3398Y}. This is faster than many forward models, because the expression for a line-of-sight integral of the product of an exponential and a sinusoid can be written analytically. A disadvantage of this method is one common to Fourier analysis: a ringing caused by the abrupt upper and lower boundaries. Ringing can be minimized by applying tapering to the perturbation. A related technique, windowing, can extract information on how the wave spectrum varies with altitude. Both windowing and tapering suggest an approach based on limiting the spatial extent of the modeled perturbation.

In this paper, we present a variation of the forward modeling method of occultation data analysis based on wavelet decomposition of the refractivity. In this method, the refractivity is assumed to  follow an exponential, with a perturbation term that is a sum of wavelets. This has the speed advantages of the Fourier decomposition method while avoiding much of the ringing. Because any refractivity profile can be decomposed into a sum of wavelets, this method maintains the flexibility and high spatial resolution of the inversion method. Additionally, it maintains the advantages of other forward-modeling methods, including the ability to handle ray crossing, diffraction, and finite source sizes, as well as the ability to calculate analytic thermal gradients while providing more more tractable statistics. 

Moreover, if we assume that the atmosphere must be statically stable to convective overturning, then we can place limits on the amplitudes of atmospheric waves. For each
vertical wavelength ($L_z$), there is a maximum stable temperature fluctuation.
This means, for each vertical wavelength, we can also find the maximum stable fluctuations in bending angle and the bending angle derivative.  This leads to four limits:  vertical wavelengths that can cause any spike (local maximum in a light curve);  vertical wavelengths that can cause bright spikes that exceed the unocculted flux;  vertical wavelengths that lead to ray crossing; and vertical wavelengths that lead to such large scattering of incoming rays that the light is offset by more then $L_z$.

Although developed for ground-based refractive stellar occultations, many of the ideas here can be applied to problems involving the line-of-sight integral through an atmosphere, such as radio occultations, UV occultations, or images of haze layering.


\section{Wavelets review} \label{sec:Meyer}

Just as an arbitrary function can be expressed as a sum of coefficients times sines and cosines (aka the Fourier Transform), it can be expressed as a sum of coefficients times wavelets, a class of functions that are compact in time, $t$, and frequency, $\omega$  
\citep{TorrenceCompo1998BAMS...79...61T, Farge1992AnRFM..24..395F,  kaiser1994friendly}.  This compactness makes wavelets  particularly useful for atmospheric studies of waves and turbulence.  Wavelet analysis allows us to identify individual waves and evaluate their statistical significance, or to measure the power spectrum vs.\ height.  Wavelets are as flexible as Fourier transforms, while avoiding, or at least isolating, many of the problems of ringing near the boundaries of a non-periodic function.  

A wavelet description begins with a unitless function $\psi(t)$ of the unitless ``time" variable, $t$. $\psi(t)$ is called the ``mother wavelet" because it spawns a family of shifted and scaled ``daughter wavelets."  Wavelets scale and translate in a way that preserves their shape and  total power.  
Daughter wavelets, translated by a real value $\Delta$ and scaled by a real value $s>0$, are related to the mother wavelet by 
\begin{equation}\label{E:meyer-phi_st}
    \psi(s,\Delta; t) = 
    |s|^{-1/2} \psi \left( \frac{t-\Delta}{s} \right)
\end{equation} 
The mother wavelet has $s=1$ and $\Delta=0$. Larger $s$ therefore means longer wavelength (smaller $\omega$).  Wavelets can also be shifted and scaled by discrete integer indices, which are related to $s$ and $\Delta$ by $s=2^{-j}$ and $\Delta=k2^{-j}$. In other words,  discrete wavelets can be related to continuous ones by $\psi_{j,k}(t)=\psi(2^{-j},k2^{-j}; t)$.

We use Meyer wavelets  \citep{SatoYamada1994JGR....9920623S}, defined in the unitless ``frequency" domain, $\omega$: 
\begin{gather}
  \setlength{\arraycolsep}{0pt}
    h(\omega) = \biggl\{ \begin{array}{ l l }
    e^{-\frac{1}{\omega^2}}, \, \omega > 0 \\
    0 , \, \omega \leq 0
  \end{array}  \\
  g_{SY}(\omega) = 
    \frac{h \left(\frac{4\pi}{3}-\omega\right)}  
    {h\left(\omega-\frac{2\pi}{3}\right) + h\left(\frac{4\pi}{3}-\omega\right)} \\ 
\hat{\phi}(\omega) = 
    \sqrt{g_{SY}(\omega) g_{SY}(-\omega)} \\
\hat{\psi}(\omega) = 
    e^{-i \omega/2}
    \sqrt{\hat{\phi}(\omega/2)^2 - \hat{\phi}(\omega)^2} 
\end{gather}
The wavelet in the time domain is then 
\begin{equation}
\psi(t) =
    \int_{-\infty}^{\infty}e^{i\omega t}\hat{\psi}(\omega)\, \frac{d\omega}{2 \pi}    
\end{equation}
    
\begin{figure}  [ht]
    \centering
    \includegraphics[width=1.0\linewidth]{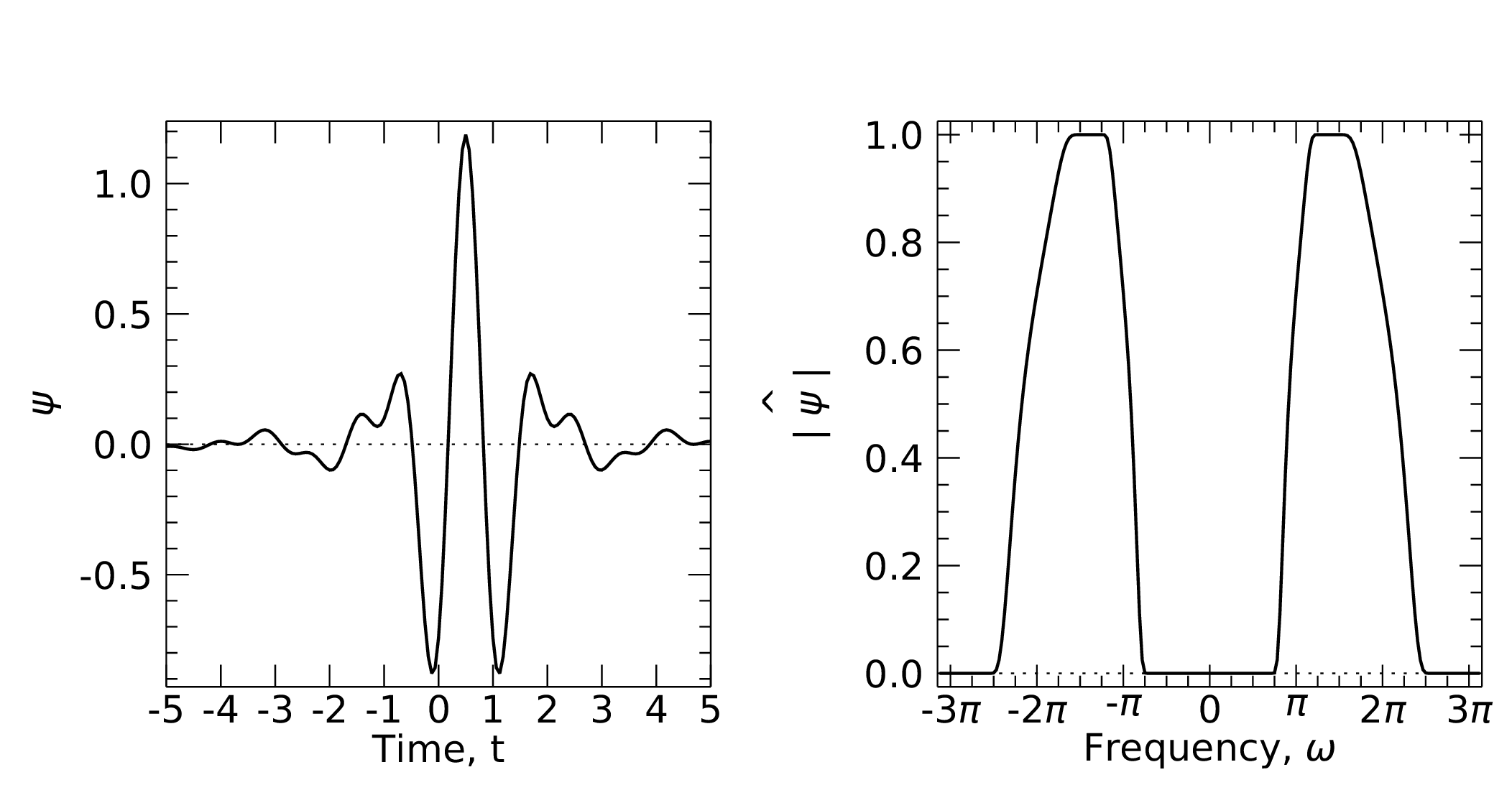}
    \caption{The mother wavelet, $\psi$ (left), and its Fourier Transform, $\hat\psi$  (right).  On the left we plot the wavelet in the time domain and on the right we plot the absolute value of the complex wavelet in the frequency domain. For the Meyer wavelet used by  \cite{SatoYamada1994JGR....9920623S} and in this work, the wavelet in the time domain is extremely smooth, and the wavelet in the frequency domain is compact, being zero except for at a small range of frequencies. }
    \label{fig:mother}
\end{figure}

The factor of $e^{-i \omega/2}$ in the definition of $\hat{\psi}(\omega)$ centers the mother wavelet
    at $t=1/2$. 
The mother wavelet is shown in Figure \ref{fig:mother}.

For Meyer wavelets, the horizontal scale of the mother wavelet is set so that shifting the wavelet by 1 in $t$ gives a wavelet that is orthogonal to the unshifted one, and for all wavelets, the amplitude of the mother wavelet is set so that the mother has unit power ($\int\psi^2dt=1$). For the version of the Meyer wavelet we use here, the power-weighted frequency for the mother wavelet is $\tilde{\omega}_\psi = 4.76 = 1.52 \pi$, and the amplitude of the mother wavelet, $A_\psi = \max{|\psi(t)|}$, is 1.19. 

Meyer wavelets have a number of useful mathematical properties.  The discrete set of $\psi_{j,k}$ forms an orthonormal set of functions. $\psi(t)$ is extremely smooth, as it is infinitely differentiable, and $\hat\psi(\omega)$ is extremely compact spectrally, being zero except for a narrow range of frequencies.  Additionally, these wavelets have been used by other authors to investigate gravity wave activity in the terrestrial atmosphere  \citep{SatoYamada1994JGR....9920623S, 
Yamada+Ohkitani_1990PThPh..83..819Y, Yamada+Ohkitani_1991FlDyR...8..101Y}. Finally, they have an infinite number of vanishing moments, making them well suited for measuring the slope of a tail of a distribution \citep{Perrier1995JMP....36.1506P, Abry1995}, which has been used to characterize turbulence or breaking gravity waves.  

\section{Light curves from an Exponential Baseline Atmosphere} \label{sec:exp+Meyerocc-decomp}

From the refractivity profile of an atmosphere, $\nu(r)$ (that is, refractivity vs.\ radius, $r$, from the planet center),  gravitational acceleration, $g$, molecular mass, $\mu$,  molecular refractivity, $K$, and  body-observer distance, $D$, we can calculate the atmospheric structure (number density, $n$, density, $\rho$, pressure, $p$, temperature, $T$, and temperature gradient, $T_r$) and models of ground-based refractive occultations (line-of-sight integral of refractivity, $\alpha$, bending angle, $\theta$, shadow radius, $y$, bending angle gradient, $\theta_r$, and normalized stellar flux, $f$). For derivations, details, citations, and further background, see \cite{Saunders2024PhDT........24Y, Sicardy2022CRPHYS....23.213S, Young2009AJ....137.3398Y, ElliotYoung1992AJ....103..991E,EshlemanGurrola1993Icar..105..298E,WassermanVeverka1973Icar...20..322W}, \cite{FrenchLovelace1983Icar...56..122F} and the foundational \cite{BaumCode1953AJ.....58..108B}. These equations assume spherical symmetry and a small bending angle.  For extensions to oblate planets or large bending angles, see \cite{Schinder+20152015RaSc...50..712S}. See Table \ref{tab:symtab} for a list of symbols. 

We write the refractivity as a baseline times a perturbation \citep{Young2009AJ....137.3398Y}. In this work, we assume an exponential baseline. That is, the atmosphere has a constant scale height, $H_0$, and the scale heights of pressure, density, and refractivity are all equal and constant with altitude \citep{BaumCode1953AJ.....58..108B}. In other words, the temperature and composition are constant and we can ignore the variation of gravity with height (valid if $r\gg H_0$, called the ``large planet approximation" by \cite{ElliotYoung1992AJ....103..991E} and the ``small scale height approximation" by \cite{Sicardy2022CRPHYS....23.213S}).
In that case, the hydrostatic altitude, $z(r)$, above a reference radius, $r_0$, is simply $z(r)=r-r_0$, and the baseline refractivity is $\bar{\nu}(r) = \nu_0 e^{-z(r)/H_0}$. (The overbar indicates the baseline atmosphere. See Table \ref{tab:acctab} for a list of accents.) In this work, we take  $r_0$ to be the radius of half-light in the baseline atmosphere, $\bar{f}_{cyl}(r_0)= {1}/{2,}$ where $\bar{f}_{cyl}$ is the flux ignoring the horizontal refocusing of starlight and far-limb effects \citep{ElliotYoung1992AJ....103..991E}.  Because $z$ is a monotonic function of $r$, we simply write $z$ for $z(r)$ and $z'$ for $z(r')$. 

The refractivity perturbation is one plus a sum of coefficients times a discrete number of unitless fluctuation functions, $c_{j,k} \, \psi_{j,k}^{\nu}(z)$, where $j$ is an index over vertical wavenumber, and $k$ is an index over altitude.  The superscript $\nu$ indicates fluctuation functions used to decompose refractivity. 

\begin{equation}\label{E:nudecomp}
\nu(r) = \bar{\nu}(r) \bigg[ 1 + \sum_{j,k}c_{j,k}\psi_{j,k}^{\nu}(z) \bigg]
\end{equation}

We have several goals in this section.  One is to relate the fluctuation functions in other parameters ($\psi_{j,k}^{T}$, $\psi_{j,k}^{\theta}$, etc.) to $\psi_{j,k}^{\nu}$. These relations are independent of the choice of $\psi^{\nu}$, and we simply write $\psi^{\nu}$ for the fluctuation function of $\nu$ in Table \ref{tab:psi}. Our second goal is to present the limits imposed by static stability on a single Meyer wavelet. We express these as the maximum fluctuation amplitude 

\begin{equation}\label{E:A}
A^\nu(s) \equiv \max|c_{crit}(s)\psi^\nu(s,0;z)|.
\end{equation}
$c_{crit}$ is the largest value of $c$ for which a single wavelet is statically stable, and depends only on the scale, $s$ (or, equivalently, the wavelength or wavenumber).
Values for $A^\nu(s)$ are given in Table \ref{tab:ampl}, as are the maximum amplitudes for the other parameters.  The final goal is to recap the calculation of an occultation light curve.

  \input{tab-psi}

\subsection{Atmospheric variables} \label{ssec:atmvar}

In this subsection, we discuss fluctuations in the atmospheric parameters, $\nu$, $n$, $\rho$, $p$, $T$, and $T_r$. We present how the fluctuation function for $\nu$ relates to that of the other parameters in rows 2-6 of Table \ref{tab:psi}, and limits on their maximum amplitudes in rows 4-7 of Table \ref{tab:ampl}.

\begin{figure} [ht] 
    \centering
    \includegraphics[width=1.0\linewidth] {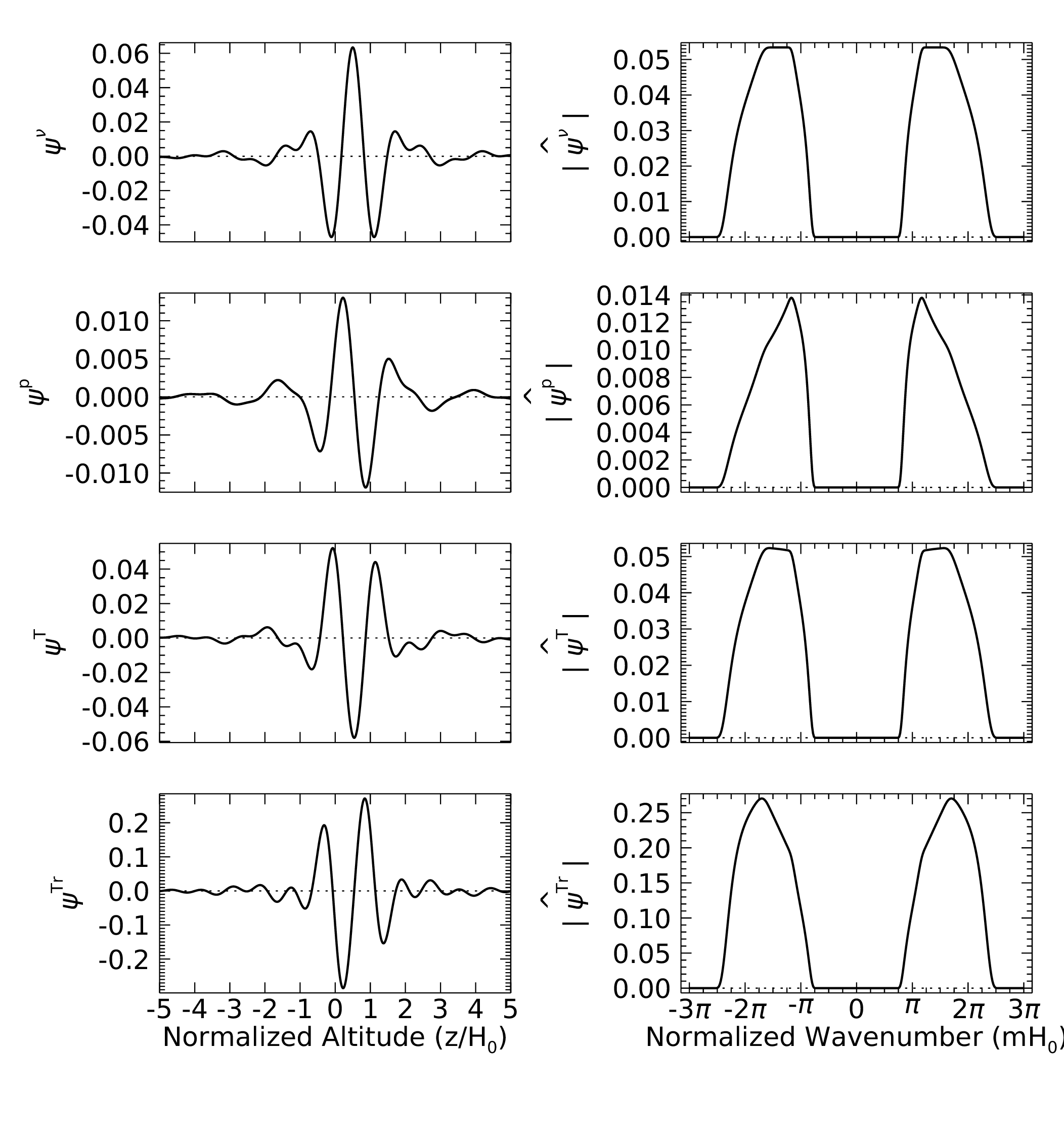}
    \caption{Statically stable fluctuations assuming Meyer wavelets for $\psi^\nu$, using $s=1$, $\Delta=0$, $c=c_{crit}=0.054$. On the left we show the fluctuations in refractivity, pressure, temperature, and temperature gradient vs.\ hydrostatic altitude in units of scale height. On the right we show the absolute value of the Fourier Transform of these fluctuations.  The fluctuations are given by $c_{crit} \psi^\nu$, $c_{crit} \psi^p$, $c_{crit} \psi^T$ and $c_{crit} \psi^{T_r}$, where $c_{crit}$ is determined by Equation \eqref{E:ATr}. This figure shows the differences in amplitude and phase of the various fluctuations, and how the changes in the Fourier Transform lead to subtle changes in shape. 
            }
    \label{fig:A01.001-atm}
\end{figure}

To use the Meyer wavelets of Section \ref{sec:Meyer}, we non-dimensionalize the hydrostatic altitude, $z$, and vertical wavenumber, $m$, with the scale height, so that $t=z/H_0$ and $\omega=m H_0$.  With this, the characteristic wavelength is $L_z/H_0=1.32 s $, where $L_z = 2\pi/\tilde{m}$, and $\tilde{m}$ is the power-weighted vertical wavenumber, $\tilde{m} H_0 =\tilde{\omega}_\psi/s$.  In this paper, we present results for  $L_z/H_0$ from $\sim 10$ to $\sim 0.001$. This range includes larger waves seen in Saturn's, Jupiter's, and Earth's mesospheres \citep{Harrington+2010ApJ...716..404H, Young+2005Icar..173..185Y, FrittsAlexander2003RvGeo..41.1003F} and smaller waves limited by Fresnel diffraction or the rate at which the occultation is observed. The Meyer wavelet is symmetric around $z/H_0=\Delta+s /2$. We plot $\psi^\nu$  for $s=1$ and $c=c_{crit}(1.0)=0.054$ in Figure \ref{fig:A01.001-atm}. It differs from Figure (\ref{fig:mother}) only by the amplitude.

The number density scales from $\nu$ via the molecular refractivity, $K$ (that is, the Lorenz-Lorentz formula for a refractive index near unity, \cite{Kragh2018Substantia....2.7K})

\begin{equation}\label{E:n}
n(r) = \frac{\nu(r) }{K} 
      = \bar{n}(r) \bigg[ 1 + \sum_{j,k}c_{j,k}\psi_{j,k}^{n}(z) \bigg]
\end{equation}
which has a particularly simple expression for the mean and fluctuation: 
$\bar{n}(r) = \bar{\nu}(r) / K = n_0 e^{-z/H_0}$, 
$\psi_{j,k}^{n}(z) = \psi_{j,k}^{\nu}(z)$, and
$\hat\psi_{j,k}^{n}(m) = \hat\psi_{j,k}^{\nu}(m)$
(See Table \ref{tab:psi}). 

Density is similarly simple, scaling the number density by the molecular mass, $\mu$, again following the  notation of \cite{Sicardy2022CRPHYS....23.213S} 
\begin{equation}\label{E:rho}
\rho(r) = \mu n(r) 
      = \bar{\rho}(r) \bigg[ 1 + \sum_{j,k}c_{j,k}\psi_{j,k}^{\rho}(z) \bigg]
\end{equation}
from which $\bar{\rho}(r) = \mu \bar{n}(r) = \rho_0 e^{-z/H_0} $,
$\psi_{j,k}^{\rho}(z) = \psi_{j,k}^{\nu}(z)$, and 
$\hat\psi_{j,k}^{\rho}(m) = \hat\psi_{j,k}^{\nu}(m)$. (See Table \ref{tab:psi}). 
Because the refractivity, number density, and density fluctuation functions are identical (Table \ref{tab:psi}), their maximum amplitudes are equal (Table \ref{tab:ampl}).

We assume hydrostatic equilibrium for the baseline atmosphere, as usual (e.g., \citealt{ElliotYoung1992AJ....103..991E}, \citealt{EshlemanGurrola1993Icar..105..298E}). In that case,
\begin{equation}\label{E:p-bar}
\bar{p}(r) = \int_{r}^{\infty}g \bar{\rho}(r') \,dr' 
\end{equation}
where $g$ is the (constant) gravitational acceleration, 
from which  
$\bar{p}(r) = g H_0 \bar{\rho}(r) = p_0 e^{-z/H_0}$. (See Table \ref{tab:psi}). This simply reproduces the ideal gas law, since $H_0=R\bar{T}/g$, where $R=k_B/\mu$ is the specific gas constant and $k_B$ is the Boltzmann constant.

Calculating $\psi_{j,k}^{p}(z)$ is a little more involved. Following \cite{Pielke2013__alma9992562341401488}, the  atmosphere is in hydrostatic equilibrium if the horizontal scales are much larger than the pressure scale height. Since  we assume spherical symmetry in this paper, horizontal scales are certainly larger than a scale height. Thus, the total pressure is also in hydrostatic equilibrium, and we have:

\begin{equation}\label{E:p}
    p(r) = 
    \int_{r}^{\infty}g \rho(r') \,dr' = 
    \bar{p}(r) \bigg[ 1 + \sum_{j,k}c_{j,k}\psi_{j,k}^{p}(z) \bigg]
\end{equation}
from which (with Equation \eqref{E:rho})
\begin{equation}\label{E:p-psi}
    \bar{p}(r) \psi_{j,k}^{p}(z) = 
   \int_{r}^{\infty}g \bar{\rho}(r') \psi_{j,k}^{\nu}(z') \,dr' 
\end{equation}

Following \cite{Young2009AJ....137.3398Y}, we then take the Fourier transform of the perturbations in $z$, and combine the Fourier and baseline exponents.  We denote the Fourier transform of a function $\psi(z)$ as $\hat{\psi}(m)$, where $m$ is the vertical wavenumber, and use the following convention for the forward and  inverse Fourier Transforms.  

\begin{equation} \label{E:FTanal}
    \hat{\psi}(m) =
    \int_{-\infty}^{\infty}e^{-imz}\psi(z)\,dz
\end{equation}

\begin{equation}\label{E:FTsynth}
    \psi(z) =
    \int_{-\infty}^{\infty}e^{imz}\hat{\psi}(m)\, \frac{dm}{2 \pi}
\end{equation}
In a real atmosphere, we cannot integrate beginning at $z=-\infty$; for solid surfaces, the atmosphere begins at the surface, and for all atmospheres, the center of the body is at $r=0$. Similarly, we cannot realistically integrate to $\infty$ because the atmosphere reaches non-collisional conditions and violates the assumption of hydrostatic equilibrium, although this limitation often has little practical effect. Fluctuations that are localized in altitude (unlike the sines and cosines in the usual Fourier Transform) allow us to choose finite integration bounds that enclose most of the power of the fluctuations.

It is useful to define a complex scale height \citep{Young2009AJ....137.3398Y}
that is independent of radius, 

\begin{equation}\label{E:H_m}
    H_m = \frac{H_0}{1-i m H_0}
\end{equation}
so that
\begin{equation}\label{E:z/Hm}
    -\frac{z'-z}{H_0} + imz'= imz -\frac{z'-z}{H_m}
\end{equation}

The pressure fluctuations are found by Equation (\ref{E:p-psi}); substituting the Fourier transform of $\psi^\nu(z')=\int e^{imz'}\hat{\psi}^\nu(m)\, {dm}/{2 \pi}$ and $\bar{\rho}(r')=\bar{\rho}(r)e^{-(z'-z)/{H_0}}$ and factoring out $g\bar{\rho}(r)$; swapping the order of integration and applying Equation (\ref{E:z/Hm}); and performing the inner integral over $r'$ (and recalling $z'-z=r'-r$)
\begin{equation}\label{E:p-psi-z}    
\begin{split}
    \bar{p}(r) \psi_{j,k}^p(r) &= 
    g\bar{\rho}(r) 
   \int_{r}^{\infty}e^{-(z'-z)/{H_0}} 
    \left[
   \int_{-\infty}^{\infty} e^{imz'}\hat{\psi}_{j,k}^\nu(m) \, \frac{dm}{2 \pi} 
   \right]\,dr'  \\ &= 
    g\bar{\rho}(r) 
    \int_{-\infty}^{\infty}
        e^{imz}
        \hat{\psi}_{j,k}^{\nu}(m)
        \left[
            \int_{r}^{\infty} 
            e^{-(z'-z)/H_m} 
        \, dr'
        \right]
    \, \frac{dm}{2 \pi} \\
    & =
    g\bar{\rho}(r) 
    \int_{-\infty}^{\infty}
        e^{imz}
        \hat{\psi}_{j,k}^{\nu}(m)
        {H_m}
    \, \frac{dm}{2 \pi}
\end{split}
\end{equation}
from which, dividing both sides by $\bar{p}(r)= g H_0 \bar{\rho}(r)$ and taking the inverse Fourier transform gives us the pressure fluctuations in terms of the refractivity fluctuations (See Table \ref{tab:psi}).
\begin{equation}\label{E:p-psi-m}    
    \hat{\psi}_{j,k}^p(m)= 
        \hat{\psi}_{j,k}^{\nu}(m)
        \frac{H_m}{H_0}  
\end{equation}

The pressure fluctuation is plotted in Figure \ref{fig:A01.001-atm}. Even for wavelengths comparable to the scale height, the pressure fluctuations are smaller than the density fluctuations, as expected. 

Temperature, from the ideal gas law, is

\begin{equation}\label{E:T}
T(r) = \frac{p(r)}{k_B n(r)}  
\approx
\bar{T} \bigg[ 1 + \sum_{j,k}c_{j,k}\psi_{j,k}^{T}(r) \bigg]
\end{equation}
In our case, $\bar{T} =  \bar{p}(r) / (k_B \bar{n}(r) )$ is constant with height.  The decomposition is only approximate because division is a non-linear operator.  For small fluctuations, we have

\begin{equation}\label{E:T-psi}
\psi_{j,k}^{T}(r) \approx - \psi_{j,k}^{\nu}(r) + \psi_{j,k}^{p}(r)
\end{equation}

Substituting Equation \eqref{E:p-psi-m} into Equation \eqref{E:T-psi} gives
\begin{equation}\label{E:T-psi-m}    
    \hat{\psi}_{j,k}^T(m) \approx 
        -\hat{\psi}_{j,k}^{\nu}(m)
        \left(
         1 - \frac{H_m}{H_0}  
        \right)
\end{equation}
(See Table \ref{tab:psi}). For fluctuations whose wavelengths are small compared to a scale height, $H_m \ll H_0$, so Equation \eqref{E:T-psi-m} reduces to the common statement that scaled temperature fluctuations are equal in amplitude and opposite in sign to scaled density fluctuations. (See Table \ref{tab:ampl} and discussion in the Atmospheric Oscillation chapter of \cite{Holton1992introduction}). 

We scale the temperature gradient, $T_r$, by the ratio of the baseline temperature to scale height, $\bar{T}/{H_0}$, and note that the baseline temperature gradient is zero in our case, $\bar{T}_r = 0$. 
\begin{equation}\label{E:dT}
T_r(r) \equiv \frac{dT(r)}{dr} 
\approx 
\frac{\bar{T}}{H_0} 
\bigg[ \sum_{j,k}c_{j,k}\psi_{j,k}^{T_r}(r) \bigg]
\end{equation}
From Equation \eqref{E:T}
\begin{equation}\label{E:dt-psi}
\psi_{j,k}^{T_r}(r) \approx {H_0} \frac{d \psi_{j,k}^{T}(r)}{dr}
\end{equation}

Because $dr/dz = 1$,  Equation \eqref{E:dt-psi} becomes $\psi_{j,k}^{T_r}(z) = {H_0} \, {d \psi_{j,k}^{T}(z)}/{dz}$. Taking the Fourier transform of both sides and bringing the derivative inside the integral for $\hat{\psi}_{j,k}^{T}$ gives:
\begin{equation}\label{E:Tr-psi-m-of-T}    
    \hat{\psi}_{j,k}^{T_r}(m) \approx 
        \hat{\psi}_{j,k}^{T}(m)
        im{H_0}
\end{equation}
which, with Equation \eqref{E:T-psi-m}
\begin{equation}\label{E:Tr-psi-m}    
    \hat{\psi}_{j,k}^{T_r}(m) \approx 
        -\hat{\psi}_{j,k}^{\nu}(m)
        im{H_0}
        \left(
         1 - \frac{H_m}{H_0}  
        \right)
\end{equation}

For a single wave to be statically stable, the total thermal gradient, $T_r(r)$, is limited by the adiabatic lapse rate, $\Gamma=g/c_p$, where $c_p$ is the specific heat at constant pressure.  Since we define $A^{T_r}$ as the maximum stable amplitude, Equation \eqref{E:dT} gives
\begin{equation}
\frac{\bar{T}}{H_0} A^{T_r} = \frac{g}{c_p}
\end{equation}
which, with the definition of $H_0$, gives
\begin{equation}\label{E:ATr}
A^{T_r} = \frac{R}{c_p} = \kappa
\end{equation}
where $\kappa$ is the symbol typically given to the ratio of the specific gas constant to the specific heat at constant pressure. $\kappa = 2/7=0.29$ for an ideal diatomic molecule.  

The amplitudes, such as $A^{T_r}$, are evaluated for an entire fluctuation, and as such are functions of the wavelet scale, $s$. For each $s$ there is a  characteristic wavelength $L_z$, and a power-weighted vertical wavenumber $\tilde{m}$.  Since many atmospheres in our solar system are  diatomic molecules (H$_2$, N$_2$), we set $A^{T_r}=2/7$ for all values of $s$ in Table \ref{tab:ampl}. We then find the value of $c_{crit}$ that produces $A^{T_r}=2/7$, and then the other wavelet amplitudes, using the relations in Table \ref{tab:psi}.
  \input{tab-ampl}

\subsection{Occultation variables} \label{ssec:occvar}

\begin{figure}[ht]\label{fig:occgeom}
    \centering
    \includegraphics[width=0.5\linewidth]{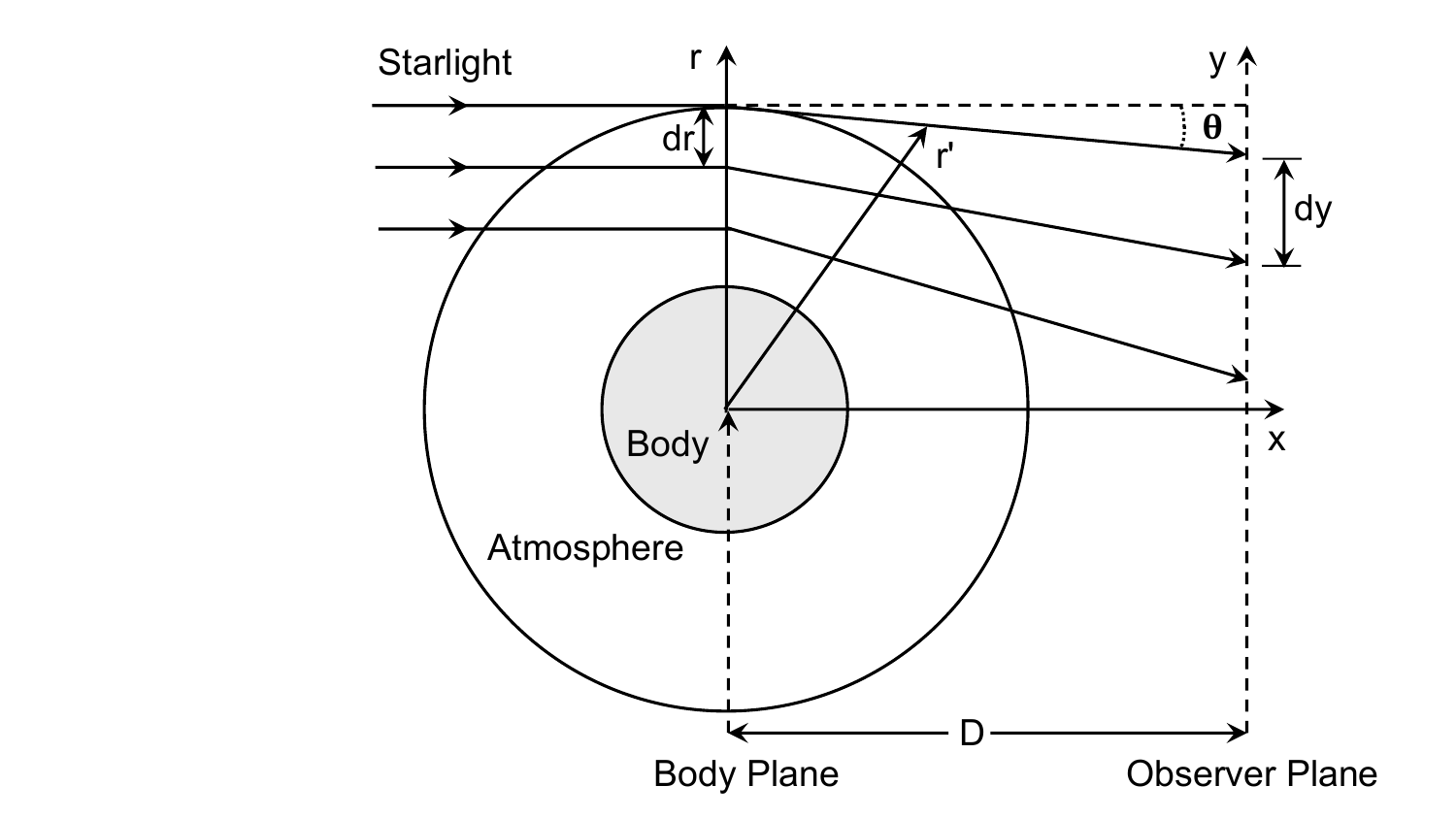}
    \caption{Refraction during a stellar occultation by a distant spherically symmetric atmosphere with a small bending angle. This schematic drawing shows the coordinates used here, where the body plane is perpendicular to the incident rays of starlight and passes through the center of the body. After \cite{ElliotYoung1992AJ....103..991E}. }
    \label{fig:F:OccGeom}
\end{figure}

The integrals for $\alpha$ (and therefore $\theta$, $\theta_r$, and  $\theta_{rr}$) require an integral along the line of sight (Figure \ref{fig:F:OccGeom}).  In this subsection, we derive their fluctuation functions in terms if $\hat{\psi}^\nu$ (see rows 7-10 of Table \ref{tab:psi}), and the limits on their maximum amplitudes (see rows 8-11 of Table \ref{tab:ampl}).

For refractive ground-based stellar occultations, we can assume negligible bending of the ray though the atmosphere, so that the radius at a line-of-sight distance $x$ on a ray with closest approach radius $r$ is given by 
\begin{equation}\label{E:r'}
    r' = \sqrt{r^2 + x^2}
\end{equation}
and the line-of-sight integral of refractivity, $\alpha$, is
\begin{equation}\label{E:alpha}
\alpha(r) = \int_{-\infty}^{\infty} \nu(r') \, dx 
      = \bar{\alpha}(r) \bigg[ 1 + \sum_ {j,k}c_{j,k}\psi_{j,k}^{\alpha}(r) \bigg]
\end{equation}
where
\begin{equation}\label{E:dx}
dx = \frac{r'\,dr'}{\sqrt{r'^2 - r^2}} 
\end{equation}
from which the baseline and fluctuations for $\alpha$ can be written
\begin{equation}\label{E:alpha-bar}    
\bar{\alpha}(r) = 
    \int_{-\infty}^{\infty} \bar{\nu}(r') \, dx 
\end{equation}

\begin{equation}\label{E:alpha-psi}    
{\bar{\alpha}(r)}  \psi_{j,k}^{\alpha}(r) = \int_{-\infty}^{\infty} \bar{\nu}(r') \psi_{j,k}^{\nu}(r') \, dx
\end{equation}

Similarly to the pressure calculation, we substitute $\bar{\nu}(r')=\bar{\nu}(r)e^{-(z'-z)/{H_0}}$ and the Fourier transform of $\psi^{\nu}$, factor out $\bar{\nu}(r)$, swap the order of integration, and apply Equation \eqref{E:z/Hm}

\begin{equation}\label{E:alpha-bar-z-lp}
\bar{\alpha}(r) = 
\bar{\nu}(r) 
    \int_{-\infty}^{\infty} 
    e^{-(z'-z)/H_0} 
    \, dx
\end{equation}

\begin{equation}\label{E:alpha-psi-z-lp}    
    \bar{\alpha}(r) \psi_{j,k}^{\alpha}(r) = 
    \bar{\nu}(r) 
    \int_{-\infty}^{\infty}
        e^{imz}
        \hat{\psi}_{j,k}^{\nu}(m)
        \left[
        \int_{-\infty}^{\infty} 
            e^{-(z'-z)/H_m} 
            \, dx
        \right]
    \, \frac{dm}{2 \pi}
\end{equation}

The next step is to perform the integral in Equation \eqref{E:alpha-bar-z-lp} and the inner integral in Equation \eqref{E:alpha-psi-z-lp}. When the radius is much larger than the scale height, we can retain only the leading terms in $H_0/r$ (aka the ``large planet" approximation \citep{ElliotYoung1992AJ....103..991E}). For the exponential atmosphere, $z'-z = r'-r$, and we can expand the exponent (e.g., \cite{BaumCode1953AJ.....58..108B}, \cite{ElliotYoung1992AJ....103..991E}) 

\begin{equation}
    z'-z = r'-r \approx 
    \frac{x^2}{2r} 
\end{equation}
with which, to first order in $H_0/r$, 
  using 
  $\int_{-\infty}^{\infty} e^{-c x^2} dx = \sqrt{{\pi}/{c}}$, for real $c>0$ or complex $c$ with real part $>0$,
we get the familiar $\bar{\alpha}(r) = \bar{\nu}(r) \sqrt{2\pi r H_0}$ 
and the less familiar 
\begin{equation}\label{E:alpha-psi-lp}    
    \bar{\alpha}(r) \psi_{j,k}^{\alpha}(r) 
    = 
    \bar{\nu}(r) 
    \int_{-\infty}^{\infty}
        e^{imz}
        \hat{\psi}_{j,k}^{\nu}(m)
        \sqrt{2\pi r H_m} 
    \, \frac{dm}{2 \pi}
\end{equation}

Dividing Equation \eqref{E:alpha-psi-lp} by  $\bar{\alpha}$ 

\begin{equation}\label{E:psi-alpha-lp}    
    \psi_{j,k}^{\alpha}(r) =     
    \int_{-\infty}^{\infty}
        e^{imz}
        \hat{\psi}_{j,k}^{\nu}(m)
        \sqrt{ 
            \frac{H_m} {H_0} }
    \, \frac{dm}{2 \pi}
\end{equation}
which, by inspection from the inverse Fourier Transform (Equation \eqref{E:FTsynth}), gives the fluctuation for the line-of-sight refractivity (Table \ref{tab:psi}).
\begin{equation}\label{E:psi-alpha-m-lp}    
    \hat{\psi}_{j,k}^{\alpha}(m) =     
        \hat{\psi}_{j,k}^{\nu}(m)
        \sqrt{ 
            \frac{H_m} {H_0} }
\end{equation}

\begin{figure} [ht] 
    \centering
    \includegraphics[width=1.0\linewidth] {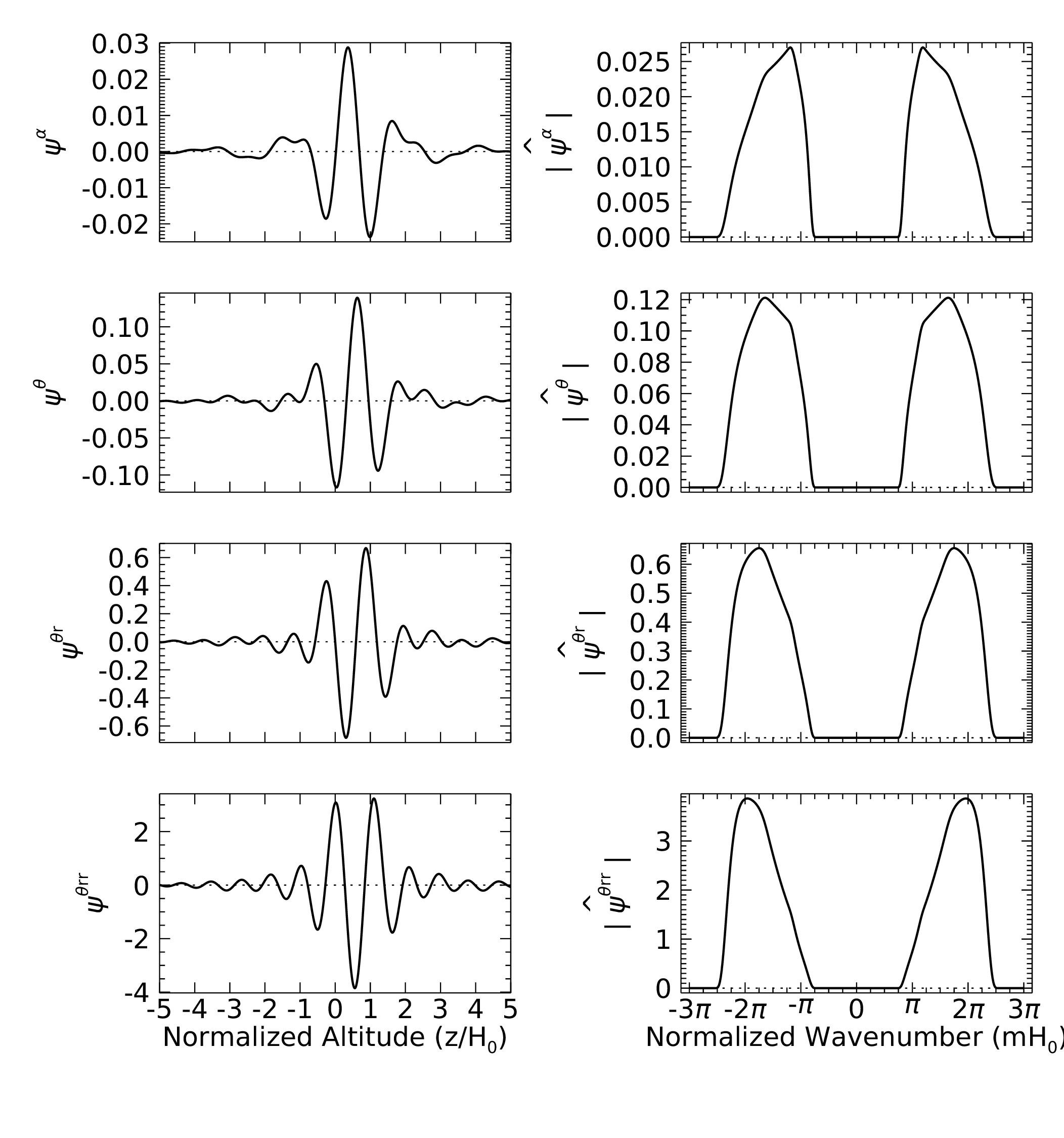}
     \caption{Statically stable fluctuations assuming Meyer wavelets for $\psi^\nu$, using $s=1$, $\Delta=0$, $c=c_{crit}=0.054$. On the left we show the fluctuations in line of sight integral of refractivity, bending angle and its derivatives vs.\ hydrostatic altitude in units of scale height. On the right we show the absolute value of the Fourier Transform of these fluctuations.  The fluctuations are given by $c_{crit} \psi^\alpha$, $c_{crit} \psi^\theta$, $c_{crit} \psi^\theta_r$ and $c_{crit} \psi^{\theta_{rr}}$, where $c_{crit}$ is determined by Equation \eqref{E:ATr}. As with Figure \ref{fig:A01.001-atm}, this figure shows the differences in amplitude and phase of the various fluctuations, and how the changes in the Fourier Transform lead to subtle changes in shape. 
}
    \label{fig:A01.001-occ}
\end{figure}
The fluctuation function $\psi^\alpha$ act like a ``half-integral" of $\psi^\nu$ (Table \ref{tab:ampl}). This supports the common observation that the line-of-sight integral averages out small scale variations.

The bending angle, $\theta$, is computed and decomposed as follows:
\begin{equation}\label{E:theta}
\theta(r)=\frac{d\alpha(r)}{dr} = \bar{\theta}(r) \bigg[ 1 + \sum_{j,k}c_{j,k}\psi_{j,k}^{\theta}(r) \bigg]
\end{equation}

To first order in $H_0/r$, the baseline bending angle is 
\begin{equation}\label{E:theta-bar=lp}\bar{\theta}(r)=d\bar{\alpha}(r)/dr=-\bar{\alpha}(r)/{H_0}\end{equation}
and the fluctuation is found from

\begin{equation}\label{E:theta-psi}    
{\bar{\theta}(r)} \psi_{j,k}^{\theta}(r) = 
 \frac
    {d \, {\bar{\alpha}(r)}  \psi_{j,k}^{\alpha}(r) }
    {dr} 
\end{equation}

We then apply the product rule; substitute Equation \eqref{E:theta-bar=lp} and $dr/dz=1$; substitute the Fourier transform of $\psi_{j,k}^{\alpha}$, factor out $\bar{\theta}$, and take the derivative inside the integral

\begin{equation}\label{E:theta-psi-lp} 
\begin{split}
    \bar{\theta}(r) \psi_{j,k}^{\theta}(r) 
    &= 
    \frac
        {d \, {\bar{\alpha}(r)}   }
        {dr} 
    \psi_{j,k}^{\alpha}(r) 
+  
    {\bar{\alpha}(r)}  \frac
    {d \, \psi_{j,k}^{\alpha}(r) }
    {dr}   
     \\
    &=
    \bar{\theta}(r)
    \psi_{j,k}^{\alpha}(r) 
-
    \bar{\theta}(r) {H_0}  
    \frac
    {d \, \psi_{j,k}^{\alpha}(r) }
    {dz}  
     \\
    &=
    \bar{\theta}(r) 
    \int_{-\infty}^{\infty}
        e^{imz}
        \hat{\psi}_{j,k}^{\alpha}(m)
        (1-{H_0}im)  
    \, \frac{dm}{2 \pi} 
    \end{split}\end{equation}

Using $(1-{H_0}im)={H_0}/{H_m}$ (Equation \eqref{E:H_m}) and Equation \eqref{E:psi-alpha-m-lp} gives (Table \ref{tab:psi})
\begin{equation}\label{E:hat-psi-theta-lp}    
    \hat{\psi}_{j,k}^{\theta}(m) 
    = 
        \hat{\psi}_{j,k}^{\nu}(m)
        \sqrt{\frac {H_0}  {H_m}} 
\end{equation}

The derivative of the bending angle with respect to radius, $\theta_r$, is similarly decomposed:
\begin{equation}\label{E:dtheta}
    \theta_r(r) \equiv \frac{d\theta(r)}{dr} 
    =
    \bar{\theta}_r(r) \bigg[ 1 + \sum_{j,k}c_{j,k}\psi_{j,k}^{\theta_r}(r) \bigg]
\end{equation}
from which

\begin{equation}\label{E:dtheta-bar}    
\bar{\theta}_r(r)
= \frac{d \, \bar{\theta}(r) }{dr} 
\end{equation}

\begin{equation}\label{E:dtheta-psi-alt}    
\bar{\theta}_r(r) \psi_{j,k}^{\theta_r}(r) 
= 
 \frac
    {d \, \bar{\theta}(r) \psi_{j,k}^{\theta}(r) }
    {dr} 
\end{equation}
which, following the same steps as for $\theta$, we find $\bar{\theta}_r(r)=-\bar{\theta}(r) /{H_0}$ and
\begin{equation}\label{E:hat-psi-dtheta-lp}    
    \hat{\psi}_{j,k}^{\theta_r}(m) 
    = 
        \hat{\psi}_{j,k}^{\nu}(m)
        \left(\frac {H_0}  {H_m}\right)^{3/2} 
\end{equation}

The derivation for the second derivative of the bending angle. $\theta_{rr}$, used in the next section, proceeds identically (See Table \ref{tab:psi}).

The fluctuation functions $\psi^{\theta}$, $\psi^{\theta_r}$, and $\psi^{\theta_{rr}}$ act like increasingly higher derivatives of $\psi^{\nu}$ (``half",``1.5$^{th}$", ``2.5$^{th}$"). As previously noted by others (e.g., \citealt{CoorayElliot2003ApJ...587L.121C}), small variations in $\nu$ can lead to large variations in $\theta_r$.

\subsection{The observed light curve} \label{ssec:lc}

With $\theta(r)$ and  $\theta_r(r)$, we can calculate the observed light curve using geometric optics. The radius from the shadow center, $y$, reached by a ray with tangent radius $r$ is
\begin{equation}\label{E:y}
y(r)=r + D \theta(r)
\end{equation}
The stellar flux diminishes as the differential refraction defocuses the starlight parallel to the refractivity gradient by $dr/dy$. If we ignore horizontal focusing and far limb contributions then we get what is called the ``cylindrical-planet" approximation to the stellar flux \citep{ElliotYoung1992AJ....103..991E}.
\begin{equation}\label{E:f_cyl}
f_{cyl}(r)=\left|\frac{1}{1 + D\theta_r(r)}\right|
\end{equation}
The starlight is also refocused perpendicular to the refractivity gradient by $r/y$, 
and the normalized stellar flux observed at $y$ from a ray with tangent radius $r$ is
\begin{equation}\label{E:f}
f(y(r))=\left| \frac{1}{1 + D\theta(r)/r}\right| \times \left|\frac{1}{1 + D\theta_r(r)}\right|
\end{equation}
As noted in \cite{ElliotYoung1992AJ....103..991E}, Equations \eqref{E:f_cyl} and \eqref{E:f} are for the normalized stellar flux from a single ray, and do not account for multiple rays striking the same location, either due to ray-crossing from over-focused rays, or from the far-limb contribution. 

They also do not account for wave-optical effects. For that, we follow \cite{FrenchLovelace1983Icar...56..122F} and \cite{goodman1996introduction} and express the flux at shadow radius $y$ as
\begin{equation}\label{E:f-wave}
f(y) = |E(y)|^2
\end{equation}
where $E(y)$ is the (complex) electric field in the observer's plane.
\begin{equation}\label{E:E-field}
E(y) = \frac{1}{\sqrt{iD\lambda}}
   \int_{-\infty}^{\infty} e^{i 2 \pi \alpha(r)/\lambda} 
                           e^{i \pi (y-r)^2/(D\lambda)} \, dr
\end{equation}
The first exponent is the phase delay at the wavelength of the observation, $\lambda$. The second exponent arises from the Fresnel approximation. The product $D\lambda$ is the square of the Fresnel scale for the unocculted star. As discussed by \cite{NarayanHubbard1988ApJ...325..503N}, diffraction needs to be considered in the case of multiple rays striking the same location, more generally termed multipath propagation.  Diffraction could add to the scintillation depending on the relative phase delays for the multiple rays, if the projected size of the star (that is, its angular diameter multiplied by $D$) is small compared to $\sqrt{D\lambda}$, and if the bandpass of the filter used to observe the occultation is narrow enough.  Stars bright enough to be observed at high time resolution and with narrow filter bandpasses also tend to have large angular diameters.  On the other hand, as the sensitivity of detectors improve, it is exciting to speculate on which occultations will be affected by wave optics.


Previous work (e.g., \citealt{Elliot+2003AJ....126.1041E}) has related, for example, the possibility of ray crossing to magnitude of refractivity fluctuations, but did not include the limits imposed by the condition of static stability (Table \ref{tab:ampl}). We can calculate occultation light curves with equations \eqref{E:f_cyl}, \eqref{E:f}, or \eqref{E:f-wave}, and we can find the fluctuations in the line-of-sight integral of refractivity and the bending angle or its derivatives from fluctuations in refractivity (Table \ref{tab:psi}). Together, this lets us use wavelets to characterize and visualize the effect of specific wavelengths at specific levels of the light curve, which we do in the following section.

\section{Implications of static stabiliy limits} \label{sec:staticstability}
We have established that, for an atmosphere to be stable, there are limits on the amplitude of the fluctuations, and that these limits depend on the vertical wavelength. In this section, we look at the implications of these limits on the morphology of light curves, and for light curve analysis. 

\subsection{Very large waves cannot make spikes } \label{ssec:nospike}

\begin{figure} [ht]
    \centering
    \includegraphics[width=0.5\linewidth]{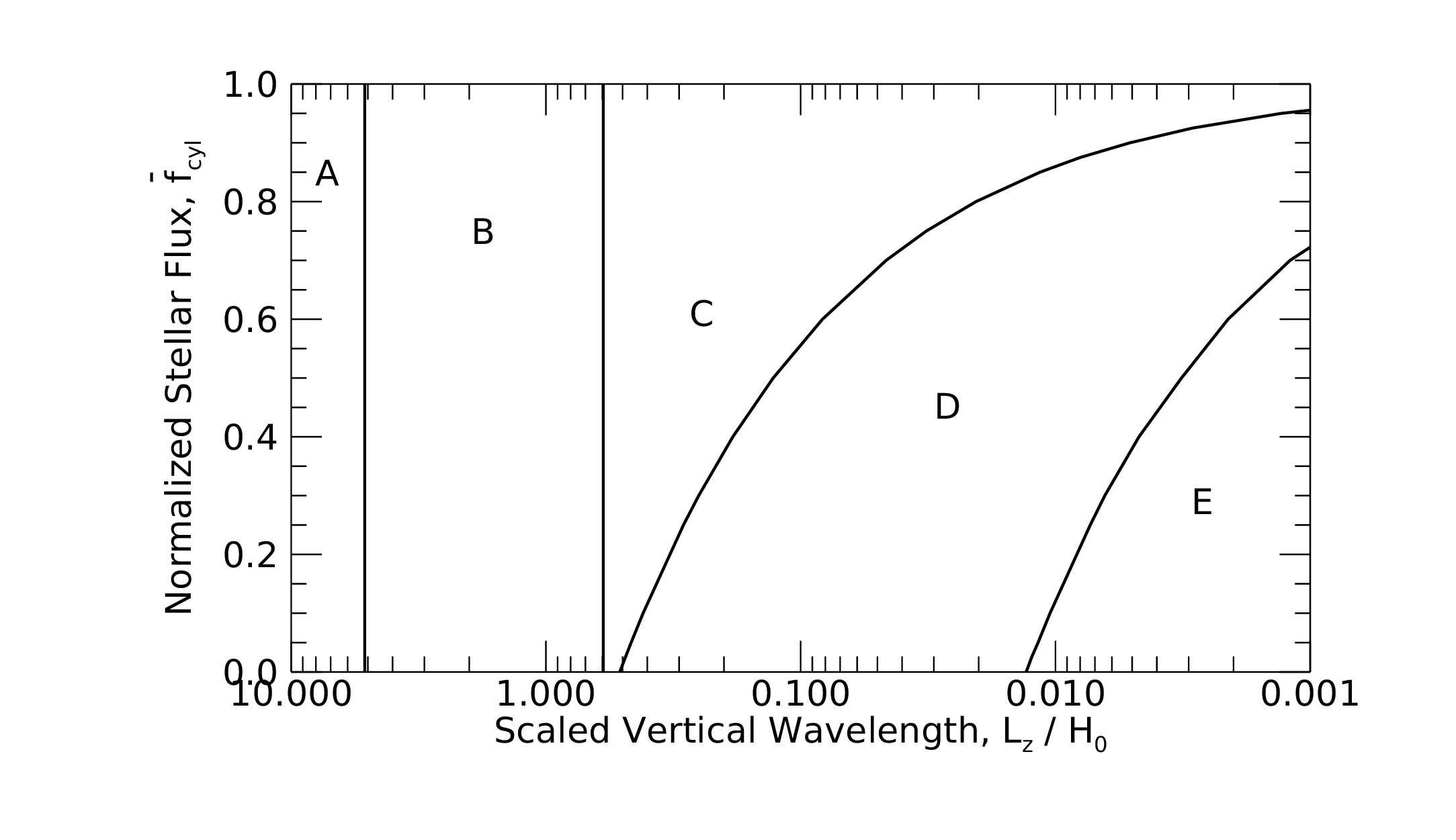}
    \caption{Light curve morphology limits.  Here we summarize the various effects discussed in Section \ref{sec:staticstability}. The baseline light curve flux level (effectively depth into the shadow) is plotted against the log of the fluctuation wavelengths.  The labeled regions (A-E) show which diagnostic regime results. For waves with wavelengths longer than 5.14 $H_0$, Region A, spikes are not possible for statically stable waves in an exponential baseline atmosphere.  Waves with vertical wavelengths longer than 0.60 $H_0$, Regions A-B, cannot make spikes brighter than the unocculted flux. Even waves with fairly small vertical wavelengths ($L_z/H_0 > 0.13$ at half-light), Regions A-C, do not lead to ray crossing.  And even waves with very small vertical wavelengths ($L_z/H_0 > 0.003$ at half-light), Regions A-D, cannot deflect the incoming starlight by more than a wavelength. }
    \label{fig:limits}
\end{figure}

Spikes are sharp increases in the flux in a light curve due to local refocusing. But waves with long vertical wavelengths are hard to distinguish from a modified background atmosphere, and cannot make spikes. What waves can make spikes? We codify that by asking what conditions are required to see any local maxima in the light curve flux. We keep our previous assumptions, that we are ignoring horizontal focusing or far-side contributions (so we are not considering central flash caustics \citep{Sicardy2022CRPHYS....23.213S}). Thus, our condition is that there is a local maximum in $f_{cyl}$ (Table \ref{tab:limits}). From 
Equation \eqref{E:f_cyl}, this occurs when the second derivative of the bending angle vanishes, or $\theta_{rr}=0$. For a single stable wavelet, this implies $1+c_{crit}\psi^{\theta_{rr}}=0$, or $A^{\theta_{rr}}=1$. From Table \ref{tab:ampl}, this condition is satisfied only for $L_z/H_0 < 5.14$ (Table \ref{tab:limits}, Figure \ref{fig:limits}). 

\begin{figure} [ht]
    \centering
    \includegraphics[width=1.0\linewidth]{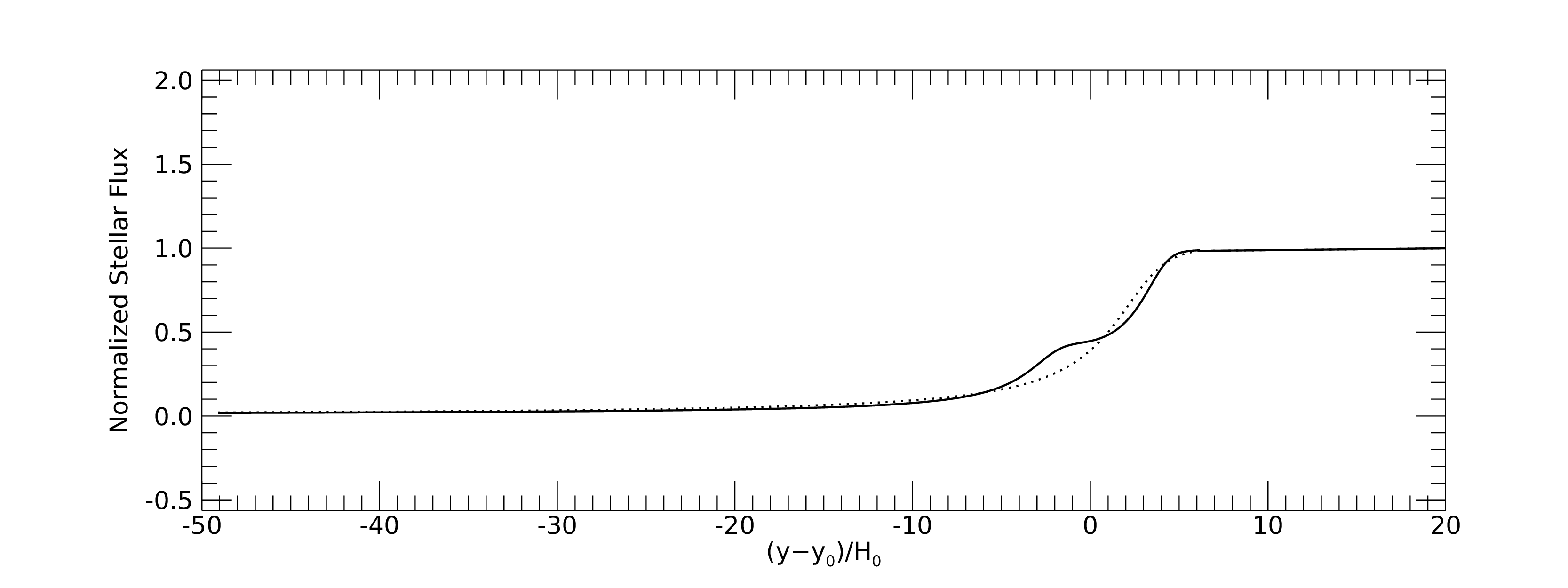}
  \caption{Simulated light curve for a statically stable wavelet at $\bar{f}_{cyl}=0.5$, with $L_z/H_0=5.14,$ and $c_{crit}=0.75$. Here we plot an occultation light curve as distorted by a single wavelet (solid) and the undistorted light curve (dotted). For this wavelet, the end of the upward fluctuation is level with the beginning of the downward fluctuation resulting in a non-negative light curve slope throughout.  A statically stable wavelet at a smaller wavelength would result in a discernible spike.
  }
  \label{fig:localmax}
\end{figure}

An example light curve is shown in Figure \ref{fig:localmax}. 
Because waves with large vertical wavelengths ($L_z/H_0 \ge 5.14 $ scale heights)
cannot make spikes, if the light curve shows spikes, or even local maxima, then there must be fluctuations with vertical wavelengths shorter than 5.14 scale heights.

\input{tab-limits}

\subsection{Moderately large waves cannot exceed the unocculted flux } 
If waves with $L_z/H_0 \leq 5.14$ are able to make spikes, the next question becomes, what waves can make  spikes that are as bright as or brighter than the unocculted flux?  That condition is $f_{cyl}=1$ (Table \ref{tab:limits}), and, from Equation \eqref{E:f_cyl}, this occurs when $\theta_{r}=0$, or $A^{\theta_{r}} = 1$. This condition is met only for $L_z/H_0 < 0.60$ (Table \ref{tab:limits}, Figure \ref{fig:limits}), for $\kappa=2/7$. 

\begin{figure} [ht]
    \centering
    \includegraphics[width=1.0\linewidth]{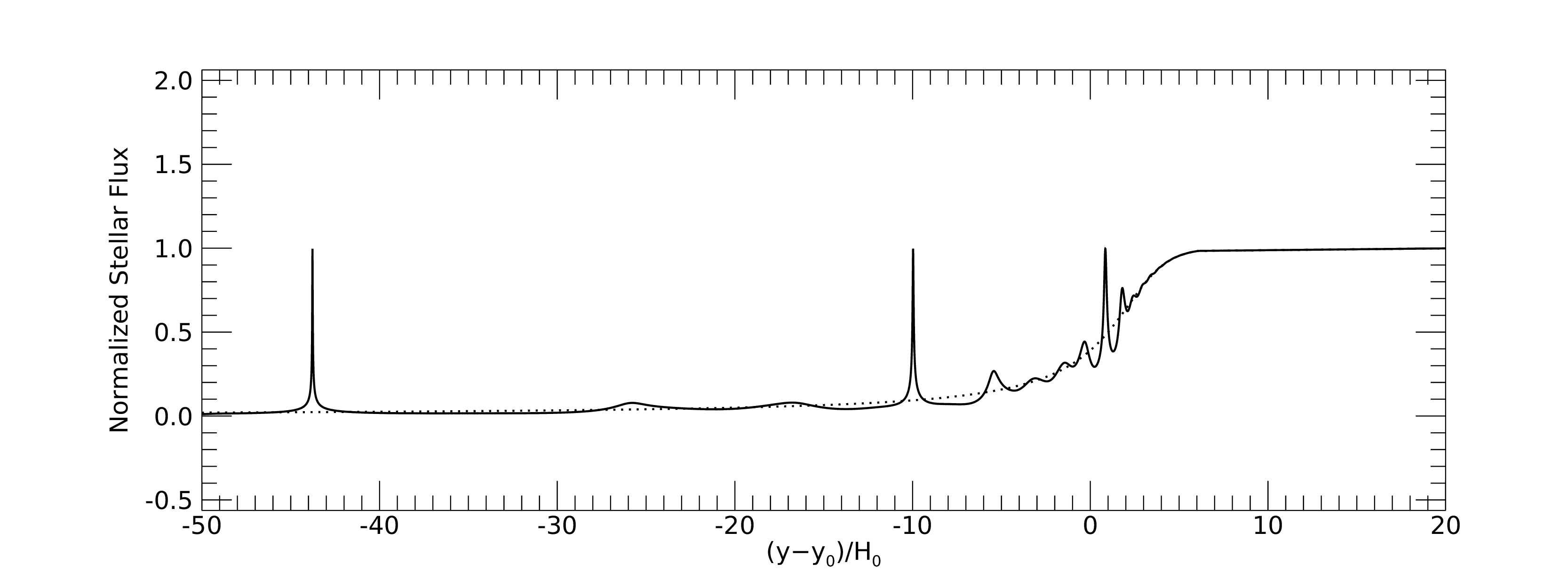}
  \caption{Occultation light curve with statically stable wavelets at $\bar{f}_{cyl}$=0.5, 0.1, and 0.025 and $L_z/H_0 = 0.60$.  We again plot an occultation light curve as distorted by waves. Note that for these values, the peak of all spikes remains at or below the unocculted baseline level.  This limit corresponds to the boundary between regions B and C in Figure \ref{fig:limits}. }
  \label{fig:fluxone}
\end{figure}

An example light curve is shown in Figure \ref{fig:fluxone}. 
Because waves with moderately large vertical wavelengths ($L_z/H_0 \ge 0.60$)
cannot make spikes brighter than the unocculted flux, if the light curve shows bright spikes, then there must be fluctuations with vertical wavelengths shorter than 0.60 scale heights. 

\subsection{Many waves cannot cause ray crossing unless they are deep into the light curve}
\begin{figure} [ht]
\centering
    \includegraphics[width=1.0\linewidth]{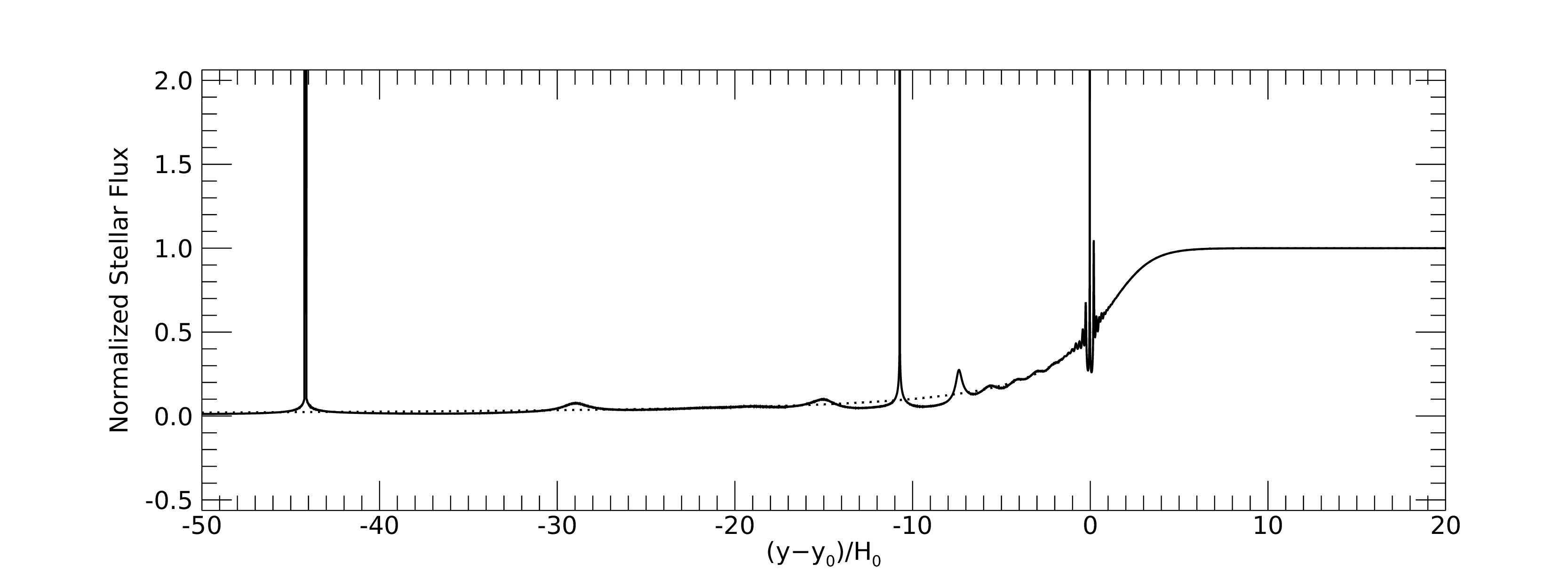}
    \caption{Light curve resulting from statically stable wavelets centered at $\bar{f}_{cyl}$=0.5, 0.1, and 0.025 and $L_z/H_0$ = 0.13, 0.42, and 0.49. Here the waves are in perfect focus (just shy of ray crossing).  In contrast to Figure \ref{fig:fluxone}, here the spikes exceed the unocculted flux level.  At amplitudes smaller than perfect focus, Abel inversions remain valid (Regions A, B, and C in Figure \ref{fig:limits}). }
  \label{fig:raycross}
\end{figure}

For waves shorter than $L_z/H_0 = 0.60$, the atmosphere is able to bring the star into sharper focus. Perfect focus happens when $dy/dr=0$ (Table \ref{tab:limits}), or, equivalently, when $1/f_{cyl}=0$ or $D \theta_r=-1$ \citep{Elliot+2003AJ....126.1041E}. Wave optics and the finite size of the occulted star prevent the flux from reaching infinity in the real world \citep{Sicardy2022CRPHYS....23.213S}.  When the star is overfocused, the atmosphere causes incoming rays to cross. In the words of \cite{Elliot+2003AJ....126.1041E} ``ray crossing occurs when the rays of starlight arrive in the observer plane in a different radial order from that in which they entered the body plane."  As discussed by \cite{Elliot+2003AJ....126.1041E}, there is a portion of the light curve, at higher flux levels, where spikes are not well-focused.  We can quantify that here.  Using the decomposition of Equation $\eqref{E:dtheta}$ and relating $D\bar{\theta}_r$ to $\bar{f}_{cyl}$, $D \theta_r=-1$ when $A^{\theta_{r}} = (1-\bar{f}_{cyl})^{-1}$. Because the wavelengths are short, we can use the approximations listed in Table \ref{tab:ampl} and find that the limiting wavelength is quadratic in $1-\bar{f}_{cyl}$ (Table \ref{tab:limits}).  At half light or above, even waves with smallish vertical wavelengths ($L_z/H_0 \ge 0.13$, for $\kappa=2/7$) cannot cause ray crossing (Fig \ref{fig:limits}).  An example light curve (Figure \ref{fig:raycross}) shows three wavelets that lead to perfect focusing.  These are calculated with geometric optics via Equation \eqref{E:f_cyl}, as are all the light curves shown in this section.

The restriction, from static stability, that some waves cannot cause ray crossing provides a simple measure of the reliability of light-curve inversion.  As light-curve inversion and the underlying Abel transformation require the linear conservation of light-curve flux, ray crossing scenarios break this fundamental assumption and render the process unreliable.  \cite{Elliot+2003AJ....126.1041E} outline various approaches to mitigate this. The first is the binning of the light curve, thus reducing the resolution in $y$; if the ray crossing occurs within a single $y$ bin, then the inversion method remains valid. This approach was improved upon by \cite{Saunders+2023PSJ.....4..199S} with the introduction of ``ratchet binning'', which attempts to make the binning process more reliable at lower flux levels. A second approach is to simply restrict ourselves to the higher flux levels of the occultation where the limiting wavelengths for ray crossing are small compared with the full resolution at which the occultation was observed (Figure \ref{fig:limits}). Here we show that we could also smooth the light curve, with a variable smoothing factor that gets coarser as the baseline flux gets smaller.  The analysis here provides the guidance for how much smoothing is required for valid Abel inversions.

\cite{Elliot+2003AJ....126.1041E} describe ways to recognize that ray crossing has occurred beyond the binning of $y$, such as caustic spikes (if the center is overfocused, there will be two locations flanking the center that are in perfect focus, and these produce spike caustics, \cite{CoorayElliot2003ApJ...587L.121C}).  While these invalidate Abel inversions, they can be examined using forward modeling, such as the wavelet technique outlined here.

\subsection{Most waves of interest do not scatter the light farther than their own wavelength } 
The above discussion about ray crossing suggests the strategy of smoothing to a chosen resolution. For example, to capture atmospheric behavior at a scale of $L_z$ or larger, for some specified $L_z$, we could smooth the light curve in the observer's plane to $L_z/\bar{f}_{cyl}$. (Because of the defocusing of starlight, the vertical scale in the observer's plane is finer than in the atmosphere plane by a factor of $\bar{f}_{cyl}$, so a vertical scale of $L_z$ in the atmosphere maps to $L_z/\bar{f}_{cyl}$ in the observer's plane.)  This smoothing strategy fails if the waves themselves scatter the incoming rays by more than their own wavelength. 

\begin{figure} [ht]
    \centering
    \includegraphics[width=1.0\linewidth]{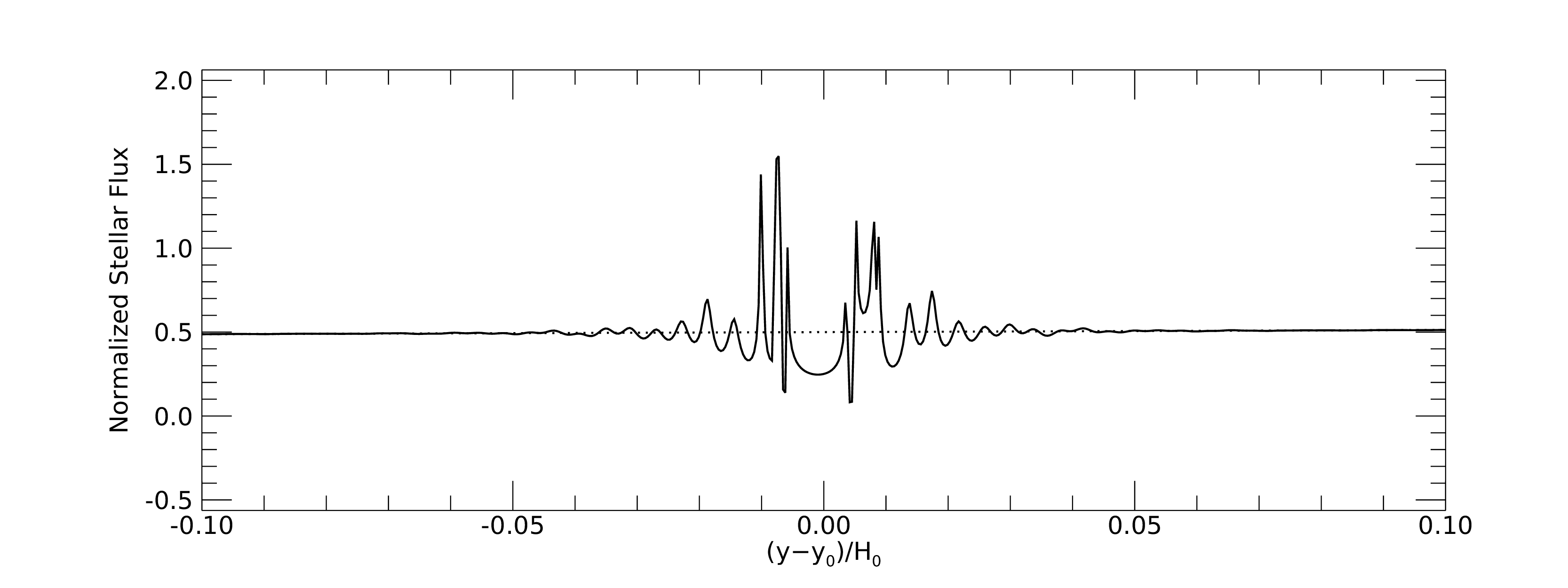}
  \caption{Simulated occultation light curve with a statically stable wavelet at $\bar{f}_{cyl}=0.5$ and $L_z/H_0 = 0.0032$. At this wavelength, the light is scattered to a distance of the characteristic wavelength around the center of the perturbation, but not beyond. Note how tightly confined all light-curve effects are at these small wavelengths. }
  \label{fig:bigoffset}
\end{figure}

The light is offset in the observer's plane by $D\theta$, and is scattered from the baseline offset by $D(\theta-\bar{\theta})$.  The limit on wavelets that scatter enough to mix signals beyond their own wavelength occurs at $D |\theta - \bar{\theta}| < L_z/\bar{f}_{cyl}$.  This can be reframed using Equation \eqref{E:f_cyl} (Table \ref{tab:limits}). Even waves with very small vertical wavelengths ($L_z/H_0 \ge 0.003$ at half light, for $\kappa=2/7$)
cannot scatter light farther than $L_z$ (Figure \ref{fig:bigoffset}). This can serve as a practical limit to avoid mixed signals.

\section{Discussion and Future Work} \label{sec:disc}

Wavelets, being compact in both altitude and wavenumber, are well suited to describing localized, quasi-monochromatic fluctuations such as turbulence or dissipating gravity waves. The derived coefficients of the wavelets can be directly related to items of physical interest, such as the power spectra of the fluctuations, with less noise than the unsmoothed Fourier transform, while also preserving altitude information. In previous atmospheric research, wavelet analysis has identified isolated dominant gravity wave modes, and followed the altitude dependence of the amplitude and wavelength of those waves \citep{Harrington+2010ApJ...716..404H, Young+2005Icar..173..185Y}. As a final advantage, wavelet analysis may be better than Fourier analysis for measuring the exponent of power spectra \citep{Abry1995}. 
This paper has laid out the foundations of wavelet analysis of occultation data, which can be extended in three  ways. 

\begin{enumerate}
        \item Arbitrary baseline atmospheres: 
        In  Section  \ref{sec:exp+Meyerocc-decomp}  we applied the general equations to the specific case of a baseline atmosphere with a constant scale height and perturbation functions based on Meyer wavelets.  Many atmospheres, such as those of Pluto, Triton, Uranus, and Jupiter, have large thermal gradients at pressures probed by stellar occultations. For these, the perturbation functions can be numerically integrated using the general equations, such as Equation \eqref{E:alpha-psi} for the line-of-sight integral. 

        \item Retrieval of temperature profiles and power spectra vs altitude with a direct occultation wavelet transform (DOWT): 
        The wavelet approach has the potential to be an alternative to Abel inversions, with well-behaved error analysis and the ability to handle ray crossing, diffraction, and finite star sizes.         
        This maintains the speed advantages of the Fourier decomposition method while avoiding some of the ringing. Because any refractivity profile can be decomposed into a sum of wavelets, this method maintains the flexibility and high spatial resolution of the inversion method. Wavelet analysis shows how a power spectrum varies with altitude. This allows, for example, not just the identification of cutoff at large $m$ (small vertical wavenumbers) indicative of the dissipation of gravity waves, but also  chance to look for the expected change in the cutoff with altitude \citep{FrenchGierasch1974JAtS...31.1707F,Smith+1987JAtS...44.1404S, SatoYamada1994JGR....9920623S, Hubbard+2009Icar..204..284H}.    
        Our suggested approach for a direct occultation wavelet transform is to retrieve wavelet coefficients at successively finer scales.

        \item Application to Uranus and other atmospheres:  
        Occultations by Uranus stop reaching the unocculted flux deep in the light curve (Figure \ref{fig:ur}).  \textbf{This my be telling us something about Uranus's atmosphere, such as a lack of waves with small wavelengths at higher pressures.  Or it may be a consequence of the fact that ray crossing can occur for larger wavelength waves than near half-light, and an over-focused fluctuation will lower the peak flux.  This can be investigated with detailed simulations of spectra of waves.}
        
\end{enumerate}


\section{Symbols and accents} \label{sec:symbols}
See Tables \ref{tab:symtab} and \ref{tab:acctab} for symbols and accents used in this paper.
\input{tab-sym}
\input{tab-acc}

\section{Acknowledgments}
Richard French and Eliot Young provided useful conversations. Leslie Young thanks Richard Binzel and Bryan Holler because even late-career scientists sometimes need advisors. The authors would also like to thank Chanita (Wata) Tubthong for her excellence with initial testing and coding in the early stages of this work.   Bill Hubbard and Dave Hinson provided insightful reviews that improved this paper.  This work was funded by NASA grants NFDAP 80NSSC20K0563 and SSO 80NSSC20K0674 as well as a grant from the MIT Undergraduate Research Opportunities Program (UROP).

%

\vspace{5mm}





\bibliography{sample631}{}
\bibliographystyle{aasjournal}



\end{document}

%% file: tab-psi.tex
\begin{deluxetable}{lccc}[htb]
\caption{Summary of Baseline and Fluctuation for an Exponential Atmosphere} \label{tab:psi} 
\tablehead{
    \colhead{Parameter} & 
    \colhead{Baseline} & 
    \colhead{Fluctuation} & 
}
\startdata
$\nu$ 
        & 
     $\bar{\nu}(r)=\nu_0 e^{-z/H_0}$
        &
    $\hat{\psi}^{\nu}(m)$
     \\
$n$ 
        & 
    $\bar{n}(r)=\bar{\nu}(r)/K$
        & 
    $\hat{\psi}^n(m)= 
        \hat{\psi}^{\nu}(m) $
     \\
$\rho$ 
        & 
    $\bar{\rho}(r)=\mu\bar{n}(r)$
        & 
    $\hat{\psi}^\rho(m)= 
        \hat{\psi}^{\nu}(m) $
     \\
$p$ 
        & 
    $\bar{p}(r)=g{H_0}\bar{\rho}(r)$
        & 
    $\hat{\psi}^p(m)= 
        \hat{\psi}^{\nu}(m)
        \frac{H_m}{H_0} $
     \\
$T$ 
        & 
    $\bar{T}(r) = \bar{p}(r)/(k_B \bar{n}(r))$
        & 
    $\hat{\psi}^T(m) \approx
        -\hat{\psi}^{\nu}(m)
        \left(1-\frac{H_m}{H_0}\right) $
    \\
$T_r$ 
        & 
    $\bar{T_r}(r) = 0$
        & 
    $\hat{\psi}^{T_r}(m) \approx
    -\hat{\psi}^{\nu}(m)
        imH_0\left( 1-\frac{H_m}{H_0} \right) $       
\\
$\alpha$ 
        & 
    $\bar{\alpha}(r) = \bar{\nu}(r)\sqrt{2 \pi r H_0}$
        & 
    $\hat{\psi}^\alpha(m) =
        \hat{\psi}^{\nu}(m)
        \sqrt{\frac{H_m}{H_0}} $ \\
$\theta$ 
        & 
    $\bar{\theta}(r) = -\bar{\nu}(r)\sqrt{\frac{2 \pi r} {H_0}}$
        & 
    $\hat{\psi}^\theta(m) =
        \hat{\psi}^{\nu}(m)
        \sqrt{\frac{H_0}{H_m}} $ \\
$\theta_r$ 
        & 
    $\bar{\theta}_r(r) = \bar{\nu}(r)\sqrt{\frac{2 \pi r} {H_0^3}}$
        & 
    $\hat{\psi}^{\theta_r} =
        \hat{\psi}^{\nu}(m)
        \left(\frac {H_0}  {H_m }\right)^{3/2} $  \\
$\theta_{rr}$ 
        & 
    $\bar{\theta}_{rr}(r) = \bar{\nu}(r)\sqrt{\frac{2 \pi r} {H_0^5}}$
        & 
    $\hat{\psi}^{\theta_{rr}}(m) =
        \hat{\psi}^{\nu}(m)
        \left(\frac {H_0}  {H_m }\right)^{5/2} $
\enddata
\end{deluxetable}

%% file: tab-ampl.tex
\begin{deluxetable}{lccccccc}[htb]
\caption{Maximum Fluctuation Amplitudes for Solitary Stable Wavelets}\label{tab:ampl} 
\tablehead{
    \colhead{wavelet scale, $s$} & 
    \colhead{   10.0} &
    \colhead{   3.89} &
    \colhead{   2.15} &
    \colhead{  0.455} &
    \colhead{  0.100} &
    \colhead{ 0.0100} &
    \colhead{0.00100} }
\startdata
$L_z/H_0$  &    13.2  &   5.14  &   2.84  &  0.600  &  0.132  & 0.0132  &0.00132\\
$\tilde{m} H_0$  &  0.476  &  1.22  &  2.21  &  10.47  &  47.60  &  476.0  &  4760.\\
$c_{crit}\approx\sqrt{\Tilde{m}_\psi}(\kappa/A_\psi)(\Tilde{m}H_0)^{-3/2}$  &     1.93  &   0.50  &   0.18  &  0.016  & 0.0017  &5.3E-05  &1.7E-06\\
$A^\nu=A^n=A^\rho\approx\kappa(\Tilde{m}H_0)^{-1}$  & 0.72  &0.30  &0.15  &0.028  &0.0062  &0.00062  &6.2E-05\\
$A^p\approx\kappa(\Tilde{m}H_0)^{-2}$  & 0.64  &0.19  &0.061  &0.0026  &0.00013  &1.3E-06  &1.3E-08\\
$A^T\approx\kappa(\Tilde{m}H_0)^{-1}$  & 0.44  &0.25  &0.13  &0.027  &0.0062  &0.00062  &6.2E-05\\
$A^{T_r}=\kappa$  & 0.29  &0.29  &0.29  &0.29  &0.29  &0.29  &0.29\\
$A^{\alpha}\approx\kappa(\Tilde{m}H_0)^{-3/2}$  & 0.68  &0.24  &0.095  &0.0087  &0.00090  &2.8E-05  &9.0E-07\\
$A^{\theta}\approx\kappa(\Tilde{m}H_0)^{-1/2}$  & 0.76  &0.38  &0.23  &0.091  &0.043  &0.013  &0.0043\\
$A^{\theta_r}\approx\kappa(\Tilde{m}H_0)^{1/2}$  & 0.85  &0.61  &0.57  &1.0  &2.1  &6.8  &21.\\
$A^{\theta_{rr}}\approx\kappa(\Tilde{m}H_0)^{3/2}$  & 0.95  &1.0  &1.5  &12.  &1.1E+02  &3.6E+03  &1.1E+05\\
\enddata
\end{deluxetable}

%% file: tab-limits.tex
\begin{deluxetable}{lccc}[htb]
\caption{Wavelength Limits on Lightcurve Morphology}\label{tab:limits} 
\tablehead{
    \colhead{Morphology} & 
    \colhead{Condition} &
    \colhead{Limits} &
    \colhead{Analytic Approximation}
}
\startdata
        Local maxima & $d(f_{cyl})/dr=0$ & $A^{\theta_{rr}} = 1$ &   $L_z/H_0 = 5.14$ \\
        Unit flux & $f_{cyl}=1$ & $A^{\theta_{r}} = 1$ &  $L_z/H_0 = 0.60$  \\
        Ray cross & $dy/dr = 0$ & $A^{\theta_{r}} = (1-\bar{f}_{cyl})^{-1}$ &  
        $L_z/H_0 \approx 0.013 (1-\bar{f}_{cyl})^2$\\
        Large scattering & $D |\theta - \bar{\theta}| = L_z/\bar{f}_{cyl} $ & 
            $A^{\theta} = ( L_z/H_0) (1-\bar{f}_{cyl})^{-1}$ &  
        $L_z/H_0 \approx 
              0.0032(1-\bar{f}_{cyl})^2$\\
\enddata
\end{deluxetable}

%% file: tab-sym.tex
\begin{deluxetable*}{lll}[htb]
\def\colheadl#1{\multicolumn{1}{l}{\vrule depth 6pt height 12pt width
0pt\relax#1}\ignorespaces}
\caption{Symbol Table} \label{tab:symtab} 
\tablehead{
\colheadl{Symbol} & \colheadl{Definition} & \colheadl{Units (cgs)}
}
\startdata
$A_\psi$ & Amplitude of the mother wavelet & unitless \\
$c$ & Coefficient that multiplies the fluctuation functions & unitless \\
$c_{crit}(s)$ & Largest value of $c$ for a statically stable wave & unitless \\
$D$ & Body-observer distance & cm \\
$E$ & Electric field & unitless  \\
$f(r)$ & Normalized stellar flux & unitless \\
$f_{cyl}(r)$ & Normalized stellar flux, cylindrical approximation & unitless \\
$g$ & Gravitational acceleration & cm s$^{-2}$\\
$H_m$ & Complex scale height for wavenumber $m$ & cm \\
$H_0$ & Scale height of baseline atmosphere at reference radius & cm \\
$j$ & Index over vertical wavenumber & integer\\
$k_B$ & Boltzmann constant & cm$^2$ g s$^{-2}$ K$^{-1}$\\
$k$ & Index over altitude & integer\\
$K$ & Molecular refractivity & cm$^3$ molecule$^{-1}$ \\
$L$ & Loschmidt's number & unitless \\
$L_z$ & Vertical wavelength & cm \\
$m$ & Vertical wavenumber & radian cm$^{-1}$\\
$\tilde{m}$ & Power-weighted mean vertical wavenumber & radian cm$^{-1}$\\
$n(r)$ & Number density & cm$^{-3}$ \\
$p(r)$ & Pressure & $\mu$bar \\
$s$ & Wavelet scale & unitless \\
$r$ & Radius from body center in the planet plane & cm \\
$R$   &  Specific Gas Constant & erg mole$^{-1}$ K$^{-1}$ \\ 
$t$    & Time domain independent variable (Section \ref{sec:Meyer}) & unitless \\
$T(r)$ & Temperature & K \\
$y(r)$ & Radius from body center in the shadow plane & cm \\
$x$ & Distance from the tangent point along the ray's line-of-sight & cm \\
$z(r)$ & Hydrostatic altitude, above $r_0$ & cm \\
$\alpha(r)$ & Line-of-sight integral of refractivity & cm\\
$\Delta$ & Wavelet translation & unitless \\
$\lambda$ & Wavelength of light for the observation & cm\\
$\mu$ & Molecular mass & g molecule$^{-1}$\\
$\nu(r)$ & Refractivity & unitless\\
$\theta(r)$ & Bending angle & radian\\
$\rho(r)$ & Density & g cm$^{-3}$\\
$\psi(t)$ & Mother wavelet& unitless \\
$\omega$  & Frequency domain independent variable (Section \ref{sec:Meyer}) & unitless\\
$\tilde{\omega}_\psi$ & power-weighted mean frequency for the mother wavelet & unitless\\
\\
\enddata
\end{deluxetable*}

%% file: tab-acc.tex
\begin{deluxetable}{lll}[htb]
\def\colheadl#1{\multicolumn{1}{l}{\vrule depth 6pt height 12pt width
0pt\relax#1}\ignorespaces}
\caption{Table of Accents etc.} \label{tab:acctab} 
\tablehead{
\colheadl{Symbol} & \colheadl{Definition} & \colheadl{Units (cgs)}
}
\startdata
$r_0$ & Reference radius & cm\\
$\bar{\nu}(r)$, $\bar{p}(r)$,   ... & Baseline $\nu$, $p$,  ... & same as $\nu$, $p$, ...\\
$\nu_0$, $p_0$, ... & Baseline $\nu$, $p$, ... at $r_0$ & same as $\nu$, $p$,  ...\\
$T_r(r)$, $\theta_r(r)$,  ... & $r$ derivative of  $T$, $\theta$,  ... & cm$^{-1}$ times units of $T$, $\theta$,  ...\\
$\psi^{\nu}(z)$, $\psi^p(z)$,  ... & Fluctuation functions for $\nu$, $p$, ...  & unitless\\
$\hat{\psi}^\nu(m)$, $\hat{\psi}^p(m)$, ... & Fourier transform of fluctuation functions & unitless \\
$A^{\nu}(s)$, $A^p(s)$,  ... & Maximum  amplitude of $c_{crit}(s) \psi^{\nu}$, $c_{crit}(s) \psi^{p}$, ... & unitless\\
$r'$, $z'$, ... & Variable of integration & cm, same as $r$, $z$, ... \\
\enddata
\end{deluxetable}